\documentclass{elsart}

 \usepackage{epsfig}

\usepackage{amssymb}

\begin{document}

\begin{frontmatter}



\title{Folding of Proteins in Go Models with Angular Interactions}


%\author[au1,au2]{Marek Cieplak, Trinh Xuan Hoang}
\author[au1]{Marek Cieplak} and
\author[au2,au3]{Trinh Xuan Hoang}

\address[au1]{Institute of Physics, Polish Academy of Sciences,
Al. Lotnik\'ow 32/46, 02-668 Warsaw, Poland}

\address[au2]{The Abdus Salam International Centre for Theoretical Physics,
Strada Costiera 11, 34100 Trieste, Italy}

\address[au3]{Institute of Physics, NCST,
P.O. Box 429, Bo Ho, Hanoi 10000, Vietnam}

\begin{abstract}
% Text of abstract
Molecular dynamics studies of Go models of proteins with the
10-12 contact potential and the bond and dihedral angle terms
indicate statistical similarities to other Go models, e.g.
with the Lennard-Jones contact potentials. The folding times
depend on the protein size as power laws with the exponents
depending on the native structural classes.
There is no dependence of the folding times
on the relative contact order even though the folding scenarios
are governed mostly by the contact order.
\end{abstract}

\end{frontmatter}

% main text
\section{Introduction}
\label{1}
There is a general conviction that protein folding is governed
by the geometry of the protein \cite{geom,tubes,tubes1,Du}
and especially by the geometry
of its native state \cite{Unger,Plaxco,Plaxco1}.
On the other hand, it is a matter of a debate
\cite{Plaxco,Plaxco1,Thirumalai,Gutin,Koga,Zhdanov,Li,biophysical}
when it comes to deciding on what does the folding
time specifically depend on, what are the proper measures of the
protein's geometry that are relevant for the kinetics,
and is there a  dependence on the system size.
In particular, a compilation of folding times at room
temperature \cite{Plaxco1} has suggested no dependence on the
number, $N$, of aminoacids and existence of a correlation
with the so called relative contact order, $CO$. This parameter is a
normalized average length along the sequence between pairs of amino acids
that interact, or make a contact. It should be pointed out, however,
that the evidence for the dependence on CO in
the $\beta$-proteins is rather weak \cite{biophysical}. The
$\beta$-proteins provide a crucial test case because their CO
is dominated by long range contacts. Furthermore, a
definition of a characteristic folding time for a protein need not
necessarily be related to a room temperature measurement. Instead,
it may require a determination of optimal kinetic conditions
for folding \cite{Koga,biophysical,optimal} and, in particular,
of the optimal kinetic temperature, $T_{min}$.\\

In this context, it is interesting to determine what are the
theoretical predictions of the coarse-grained
Go-like \cite{Goabe,Stakada} models since their construction
depends primarily on the geometry of the native state and thus
these models should be particularly attuned to the features of geometry.
In ref. \cite{biophysical}, we have studied the scaling properties
of the folding times  of 51 proteins in two versions of the Go model.
In the current paper, we report similar results on the third model,
henceforth called model C,  which
has been used by Clementi, Nymeyer, and Onuchic \cite{Clementi}
for several proteins. It has also been used by
Koga and Takada \cite{Koga} to study folding times of many proteins at
temperatures corresponding to the maximal specific heat and to propose
a correlation with CO$\cdot N^{0.6}$, as we discuss further
in reference \cite{biophysical}.\\

Model C, in addition to the
pairwise distance-dependent interactions of the 10-12 form
and  harmonic tethering forces between subsequent amino acids,
incorporates three and four body
terms in the potential which correspond to angular
dependencies. This model has been shown to posses an explicit two-state
behavior since its free energy, at least for some proteins,
has a two-minima form when plotted against the fraction, $Q$, of
the established native contacts. Thus in the model of Clementi
et al., $Q$ is believed to be 
the reaction coordinate for folding whereas in the other
models studied in ref. \cite{biophysical} the explicit reaction
coordinate is not known. We find that all three models
show nearly identical behavior even though individual values
of the parameters may display sensitivity to the choice of the model.
In particular, there is a division
into (at least) three kinetic universality classes which agree with the
three structural classes: $\alpha$-, $\beta$-, and $\alpha - \beta$
proteins, where the division is based on what are the kinds of secondary
structures that are present in the native state. In each class, there is
a separate growing trend of the folding time with $N$. This trend
appears to be better described by a power law than by an exponential
function.\\

Another result for model C is that
there is still no correlation with CO.
It should be noted that the folding scenarios can be described
in terms of graphs that show times, $t_c$, 
needed to establish contacts plotted againts the contact lengths.
In all three models, these graphs are dominated by
a monotonically increasing function of $j-i$, where $i$ and $j$
are the locations of the amino acids involved.
A strictly monotonic dependence can be characterized by a
single parameter, like CO, but the deviations are outside of the scope
of such a description. It is the usually present deviations from
the monotonic trends that should be responsible
for the lack of correlation with CO.
An alternative point of view though has been recently proposed by
Plaxco et al. \cite{Pande} and it is that simple Go models lack
cooperativity.
They suggest that this cooperativity can be introduced by making
contact energies depend on conformation through $Q$ and studies of
$N$=27 lattice models suggest a correlation with CO.\\

% \appendix

 \section{Go models}
 \label{2}
The general basic prescription for the construction
of a Go model of a protein is to use a Hamiltonian
that incorporates the chain-like connectivity and
such that the ground state of the model agrees with the experimentally
determined native conformation. Clearly, there are many ways of
accomplishing this but they are expected to be physically similar.
Our approach is outlined
in references \cite{Hoang,Hoang1} with the updated
details given in \cite{biophysical}.
Briefly, the amino acids
are represented by point particles of mass $m$ located at the
positions of the C$^{\alpha}$ atoms. They are tethered by a strong
harmonic potential with a minimum at the peptide bond length. The
native structure of a protein is taken from the
PDB \cite{PDB} data bank  and the interactions between the
amino acids are grouped into native and non-native.
The distinction is based on taking the fully atomic representation
of the amino acids in the native state and then checking for overlaps
assuming the van der Waals radii of the atoms. The procedure
\cite{Tsai,prion} involves multiplication of the radii by 1.24
to take softness of the potential into account.
The amino acids ($i$ and $j$) that are found to overlap in this
sense are considered to be forming contacts and the corresponding
contact range is from about 4.4 to 12.8 $\AA$. These pairs are
endowed with a pairwise attractive potential, $V_{ij}$,
between the $C^{\alpha}$
atoms such that its minimum agrees with the experimentally
determined distance in the native state.
The non-native contacts are purely repulsive
and corespond to a core of radius $\sigma = 5 \AA$.\\

The choice of the attractive native potential and a selection of
additional terms in the Hamiltonian is what makes a distinction between
various versions of the model.  In model A
\begin{equation}
V_{ij} =
4\epsilon \left[ \left( \frac{\sigma_{ij}}{r_{ij}}
\right)^{12}-\left(\frac{\sigma_{ij}}{r_{ij}}\right)^6\right],
\end{equation}
whereas in models B and C
\begin{equation}
V_{ij} =
\epsilon \left[ \left( 5\frac{r^{(n)}_{ij}}{r_{ij}}
\right)^{12} - 6 \left(\frac{r^{(n)}_{ij}}{r_{ij}}\right)^{10}\right],
\end{equation}
where $r^{(n)}_{ij}$ coincides with the native distance.
This potential is believed to correspond to the hydrogen bonds
better than the Lennard-Jones one. $\epsilon$ is the uniform scale
of the energy.
In models A and B the additional terms correspond to four body forces
that favor the native sense of chirality \cite{biophysical}.
Model C is our primary focus in this paper and instead of the
chirality potential it is equipped with
the angular terms:
\begin{equation}
V^{BA} = \sum_{i=1}^{N-2} K_\theta (\theta_i - \theta_{0i})^2
\end{equation}
\begin{equation}
V^{DA} = \sum_{i=1}^{N-3} [ K^1_{\phi}(1+cos(\phi_i - \phi_{0i})+
K^3_{\phi}(1+cos3(\phi_i -\phi_{0i}))]
\end{equation}
where $\theta _i$ and $\phi _i$ represent the bond and dihedral
angles respectively and the subscript 0 indicates the native values.
The bond angle is determined by three
subsequent residues and the dihedral angle by four subsequent
residues  (by forming vector products of two subsequent
residue-to-residue position vectors). Following ref. \cite{Clementi},
we take  20$\epsilon$, $\epsilon$, and $0.5\epsilon$ for $K_{\theta}$,
$K^1_{\phi}$, and $K^3_{\phi}$ respectively and the native
contacts with $|i-j| < 4$ are discarded.\\

The ground state of the model corresponds to the native state
at room temperature.
The thermal fluctuations away form this state are accounted
for by introducing the Langevin noise \cite{Grest}
with the damping constant $\gamma$
of 2 $m/\tau$, where $\tau$ is $\sqrt{m \sigma ^2 /\epsilon}$.
This corresponds to the situation in which the inertial effects are
negligible  \cite{biophysical}
but a more realistic account of the water environment
requires $\gamma$ to be about 25 times larger \cite{Veitshans}. Thus
the folding times obtained for $\gamma$=2$m/\tau$ need to be multiplied
by 25 since there is a linear dependence
on $\gamma$ \cite{Hoang,Hoang1}.\\

 \section{Results}
 \label{3}
The list of the proteins studied is exactly the same as in ref.
\cite{biophysical}. 21, 14, and 16 are of the $\alpha - \beta$,
$\alpha$, and $\beta$ kind respectively. The temperature dependence
of the folding time was determined as a median time
corresponding to at least 101 folding processes. The median
folding time, as determined at $T_{min}$, is denoted by $t_{fold}$.
The results for model C on the log-log and log-linear scales are shown in
Figures 1 and 2 respectively. Each panel corresponds to a different
structural class. The $\alpha$- proteins exhibit the slowest
growth with $N$. Compared to  models A and B
\cite{biophysical}, the folding times for
individual proteins may be quite distinct, but the general location of the
data points and the trends are the same. Thus, statistically,
all of the three models are equivalent and the growing trends with
$N$ are the same. One way to illustrate this is shown in Figure 2
in which the folding times obtained in model C are plotted against
those in model A. The data points are clustered around the
diagonal direction and the correlation level is 85\%.\\

The power law fits (Figure 1) correspond to the
exponents of 2.5, 1.7 and 3.2 for $\alpha - \beta$, $\alpha$, and $\beta$
respectively. The exponential law fits (Figure 3) yield correlation
lengths correspondingly of 32, 42 and 24. All of these values are
consistent with what was obtained for models A and B and the fits
to the exponential laws are worse (the overall correlation level
for the power laws in model C is 81\% but 74\% for the exponential fits)
However,
in model C, the scatter of the data points around the trends
is somewhat larger than in the other two models (the corresponding sets
of numbers are 86\%, 82\% and 89\%, 87\% for models A and B
respectively).\\

Figure 4 shows the behavior of characteristic temperatures, $T_f$ and
$T_{min}$, for the 51 proteins.
The former is the folding temperature. It is determined
by the condition that the probability of staying in the native
basin is near $\frac{1}{2}$. Compared to  models A and B, the values of
the two temperatures are, roughly, twice as high but they stay
comparable to each other, indicating overall good folding
properties (in models of random sequences of amino acids, $T_f$ is
substantially lower than $T_{min}$). Similar to models A and B,
there are no growing trends with $N$ in $T_f$.
In model A, there appears to be a growing trend in $T_{min}$ for
the $\alpha$ and $\alpha - \beta$ proteins \cite{biophysical}.
In model C, on the other hand (Figure 4),
there is either no net dependence on $N$ or a weak growing trend
in $T_{min}$ for each structural class.\\

It should be noted that the twice as big characteristic temperatures
found in model C, when compared to model A, also affect a
scale of a  typical
dependence of the folding time, $t_{fold}$, on $T$.
This is illustrated for the protein 2ci2 in Figure 5.
In model C, the dependence is much broader than in model A and this is
the usual situation. There are proteins, however, like 1aho,
in which the opposite takes place and then $t_{fold}$ at $T_{min}$ in
model A is shorter (by a factor of 2.5 in this case) than in C.
There are also proteins, like crambin, for which the widths of
the U-shaped curves are about the same and so are the folding times
at $T_{min}$.\\

The dependence on CO is also very much like in models A and B.
Figure 6 shows that neither $t_{fold}$ nor the characteristic
temperatures exhibit any clear trend with CO, except maybe for a
very weak growth of $T_f$.\\

The folding scenarios, however, do depend on the contact order. This is
illustrated in Figure 7 for the case of the proteins with the PDB code of
2ci2 and for all of the  three models. Independent of the model,
there is a basic monotonic growth of the average time needed to establish
a contact for the first time. The average is obtained based on 400
folding trajectories at $T_{min}$.
Most events are governed by this
monotonic trend but there is a lower side branch which introduces
double valuedness in the dependence on $|j-i|$. Thus the contact
order cannot fully describe the folding process which explains no
dependence on CO. It is remarkable that the folding scenarios  are
so insensitive to the detailed version of the  Go model.\\

Model C studied here is nearly the same as the
one considered by Clementi et al. \cite{Clementi} but there are
two differences. The first one is that the contact maps are not
exactly the same because of the different procedures used
to determine the contacts. The second is a different molecular dynamics
scheme. Clementi et al. use a leapfrog algorithm and control
the temperature by velocity rescaling. Ours is the fifth order
predictor-corrector scheme with the Langevin noise as a thermostat.
As expected, our results for the equilibrium quantities
are very close. This is illustrated in Figure 8 which shows
the free energy as a function of $Q$ and specific heat as a function
of $T$ for the protein 2ci2, both obtained by the histogram method
\cite{Swendsen,WHAM}.
The two-state behavior and values of the quantities
are in agreement with Clementi et al. The kinetic properties, however,
need not agree due to a different account of the effective viscosity
of the environment.\\

In summary, the choice of a version of the Go model,
within the same molecular dynamics scheme, is not very relevant
in statistical studies of model proteins even though it may affect
properties of individual systems. Each Go model suggests same power law
dependence on $N$ of the folding time and no dependence on a single
average contact order. It would be interesting to study whether
the cooperativity effects, as discussed by Plaxco et al. \cite{Pande}
can indeed generate demonstrable trends as a function of CO.\\

This work was funded by KBN.

\newpage
\centerline{FIGURE CAPTIONS}

\begin{description}
\item[Fig. 1.]
The scaling of $t_{fold}$ with $N$ for the 51 proteins in model C.
The data are split into the structural classes as indicated.
The lines correspond to th epower law exponent displayed
in the right corner of each panel. The error bars on the exponent
are of order $\pm$0.2. The folding times are  calculated at $T_{min}$.
The correlation levels of the points shown are 78\%, 98\%, and 87\%
for the top, middle, and bottom panels respectively.

\item[Fig. 2.]
The log-log plot of the
folding times at $T_{min}$ for the 51 proteins as obtained
in model C versus those obtained in model A.

\item[Fig. 3.]
The data of Figure 1 redisplyed on the log-linear plane. The dashed
line indicate fits to the exponential law $t_{fold} \sim exp(b/\xi )$
with the values of $\xi$ shown in the right corner of each panel.
The correlation levels are 68\%, 97\%, and 81\% respectively.

\item[Fig. 4.]
The values of $T_{min}$ and $T_f$ shown versus $N$ for model C
for the three structural classes.
The solid lines indicate the average values. The dotted lines show
the average values of $T_{g2}$ -- the temperature at which the
median folding time is twice as long as at $T_{min}$ and $T_{g2}$
is on the lower temperature side of $T_{min}$.

\item[Fig. 5.]
The folding time, defined as the first passage time, for the
protein 2ci2 as a function of $T$ in the two models.
The solid guiding line and filled circles are for model C and the dotted
line with the open square symbols are for model A. The arrows
indicate the values of $T_f$.
The values of $T_{min}$ are 0.25 and 0.48 $\epsilon /k_B$ for models
A and C respectively.

\item[Fig. 6.]
The dependence of $t_{fold}$, $T_{min}$, and $T_f$ on the relative
contact order parameter for model C. The data symbols indicate the
structural classes and are identical to thise in Figs. 1,2, and 3.

\item[Fig. 7.]
The average times to form contacts of a given length $|j-i|$
for the first time  in the three
models. The error bars are of the order of the size of the symbols.

\item[Fig. 8.]
The thermodynamic parameter for 2ci2 in model C.
The top panel shows the specific heat (the maximum is located
at a temperature which is nearly twice as high as $T_f$).
The bottom panel shows the free energy as a function of the
fraction of established native contacts at the temperature
corresponding to the maximum in the specific heat.

\end{description}

%FIGURE 1
\begin{figure}
\epsfxsize=6in
\centerline{\epsffile{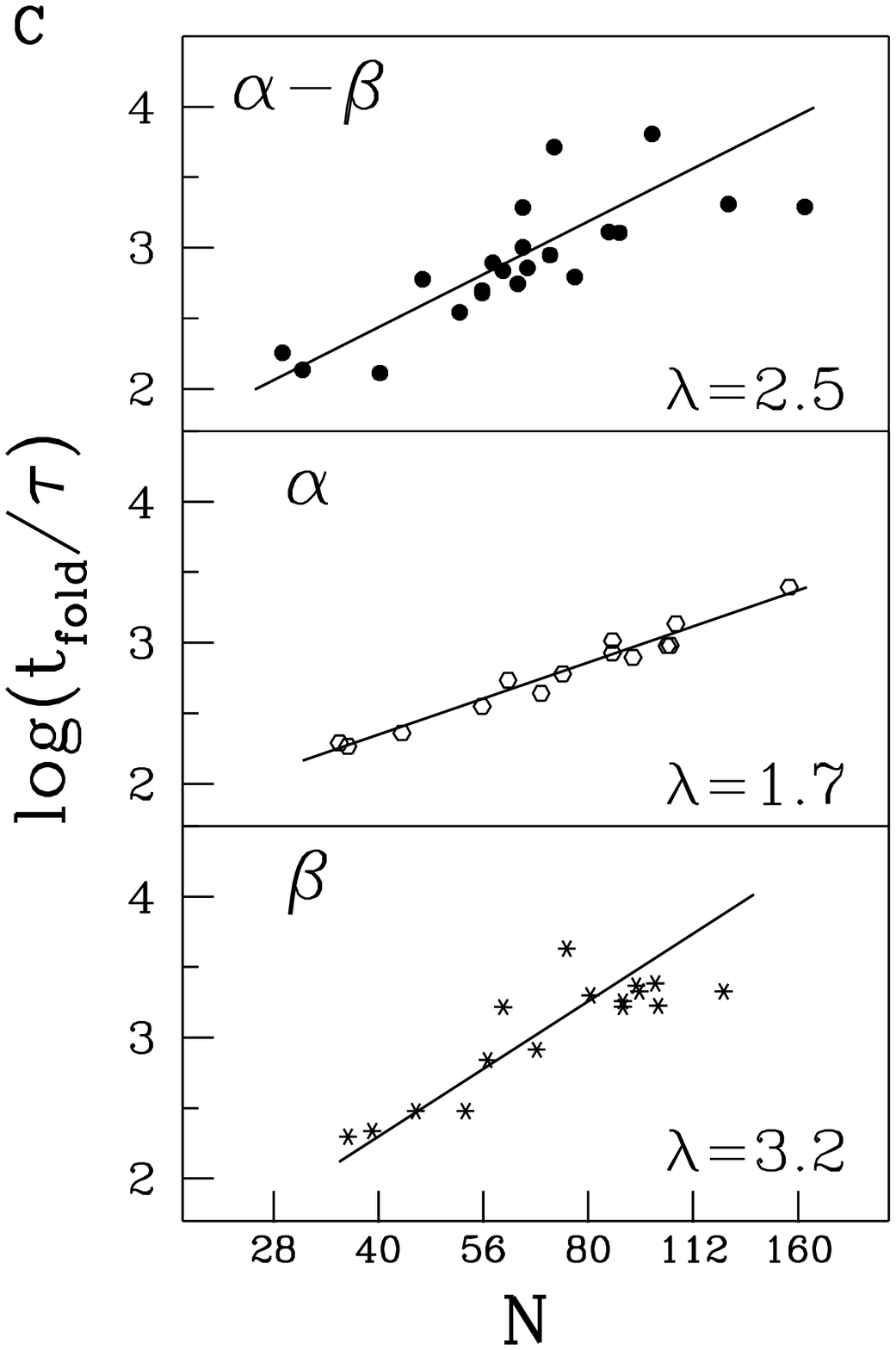}}
\caption{ }
\end{figure}

%FIGURE 2
\begin{figure}
\epsfxsize=6in
\centerline{\epsffile{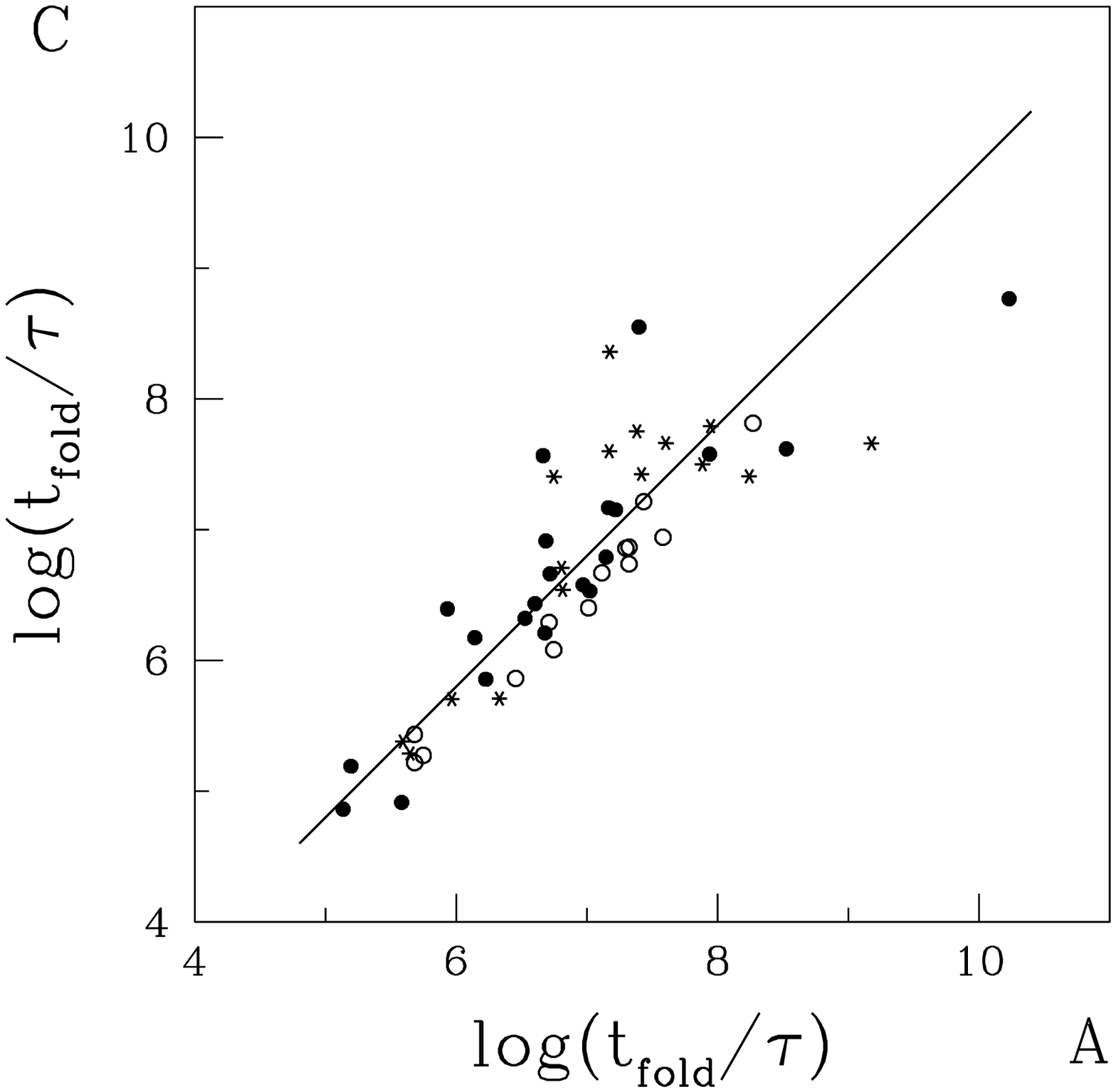}}
\caption{ }
\end{figure}

%FIGURE 3
\begin{figure}
\epsfxsize=7in
\centerline{\epsffile{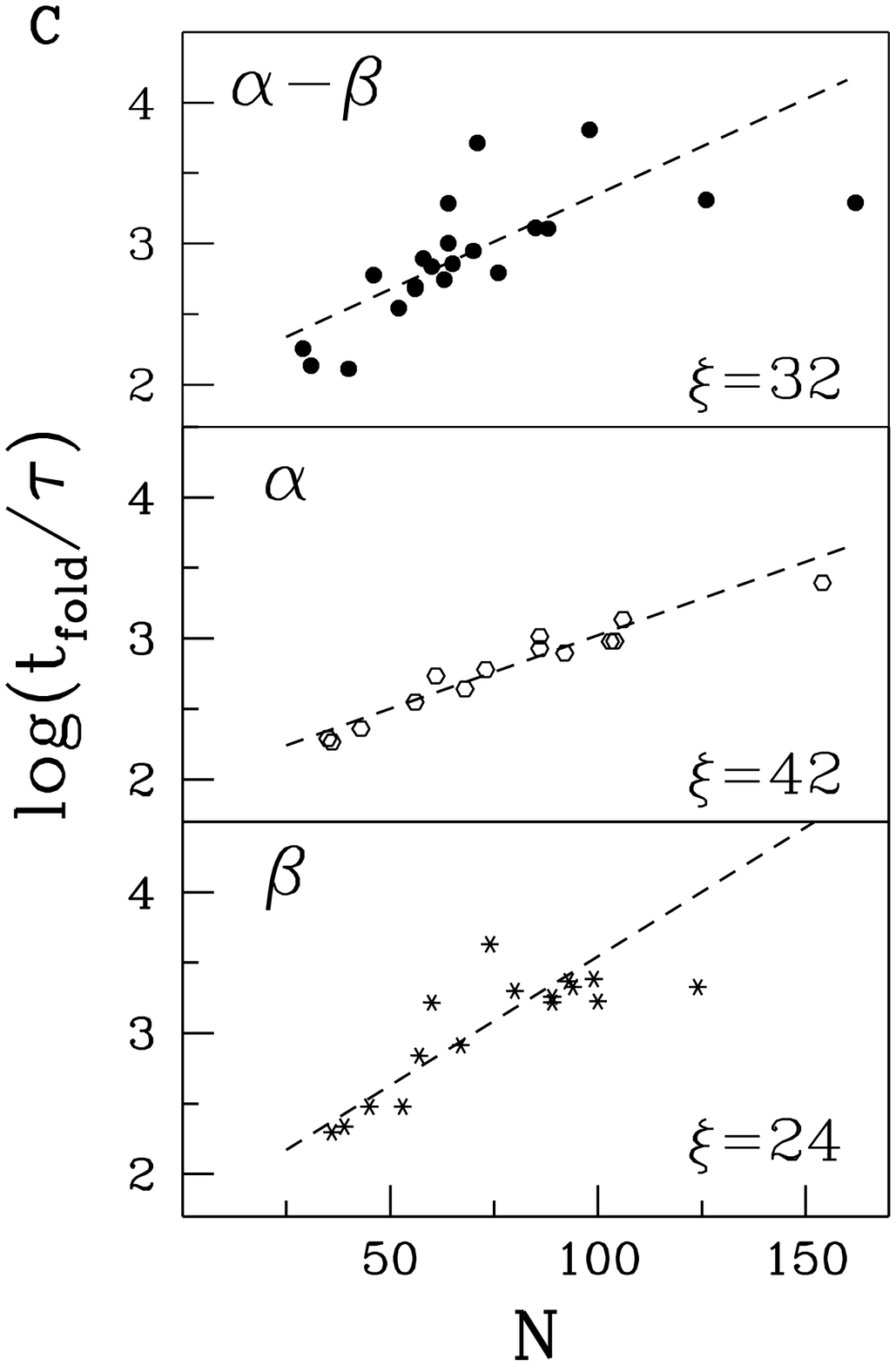}}
\vspace*{3cm}
\caption{ }
\end{figure}

%FIGURE 4
\begin{figure}
\epsfxsize=7in
\centerline{\epsffile{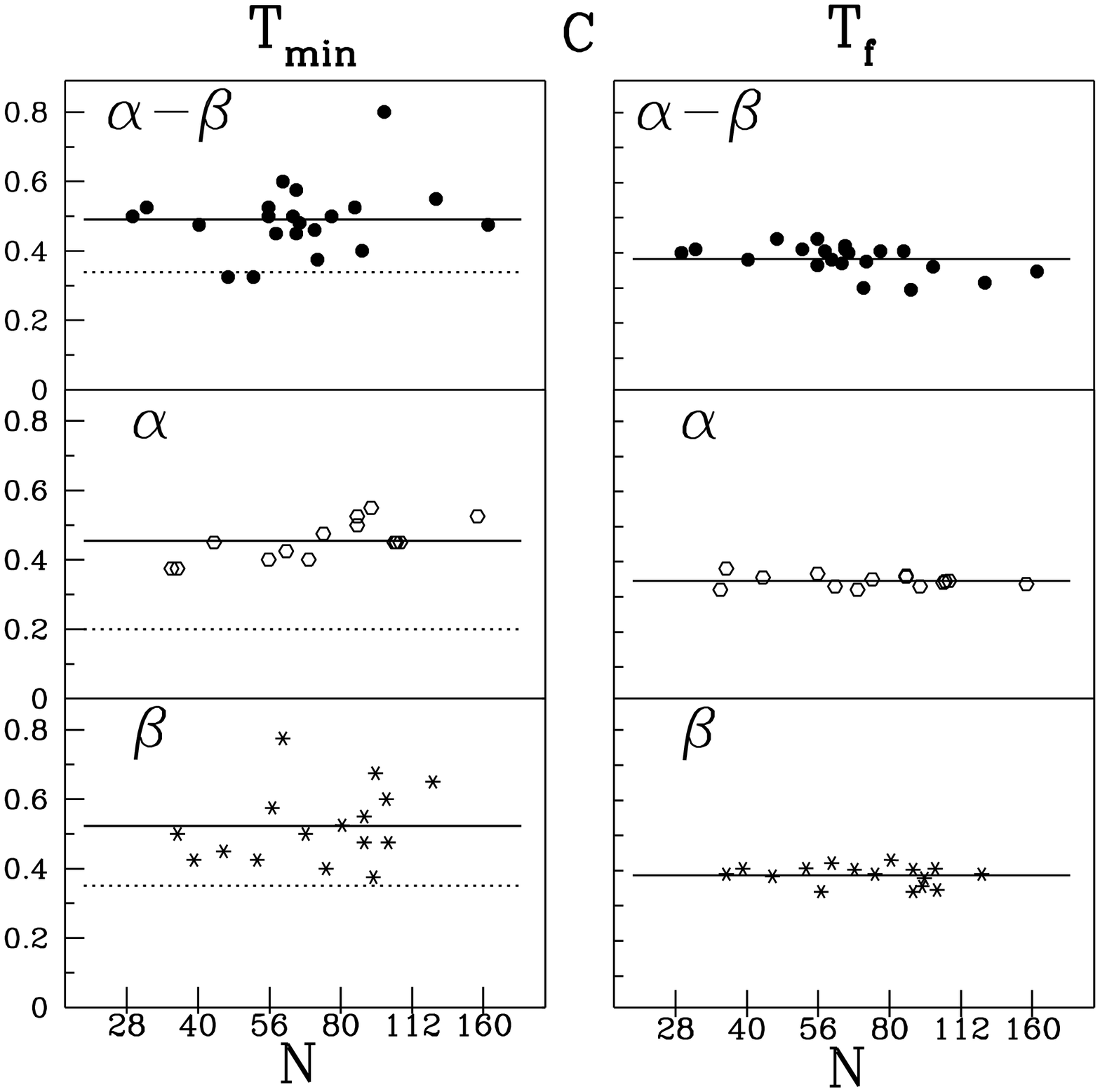}}
\vspace*{3cm}
\caption{ }
\end{figure}

%FIGURE 5
\begin{figure}
\epsfxsize=7in
\centerline{\epsffile{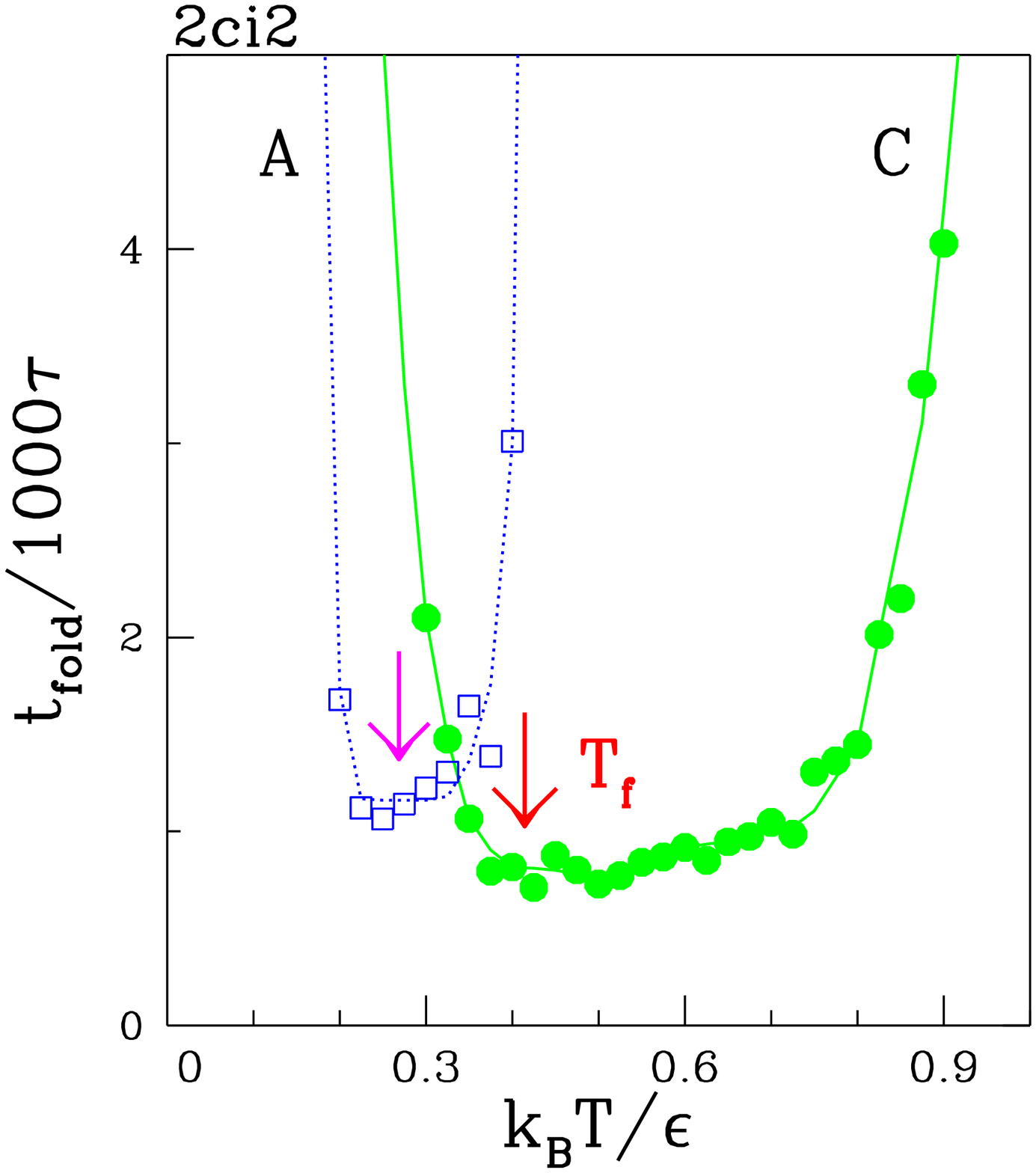}}
\vspace*{3cm}
\caption{ }
\end{figure}

%FIGURE 6
\begin{figure}
\epsfxsize=7in
\centerline{\epsffile{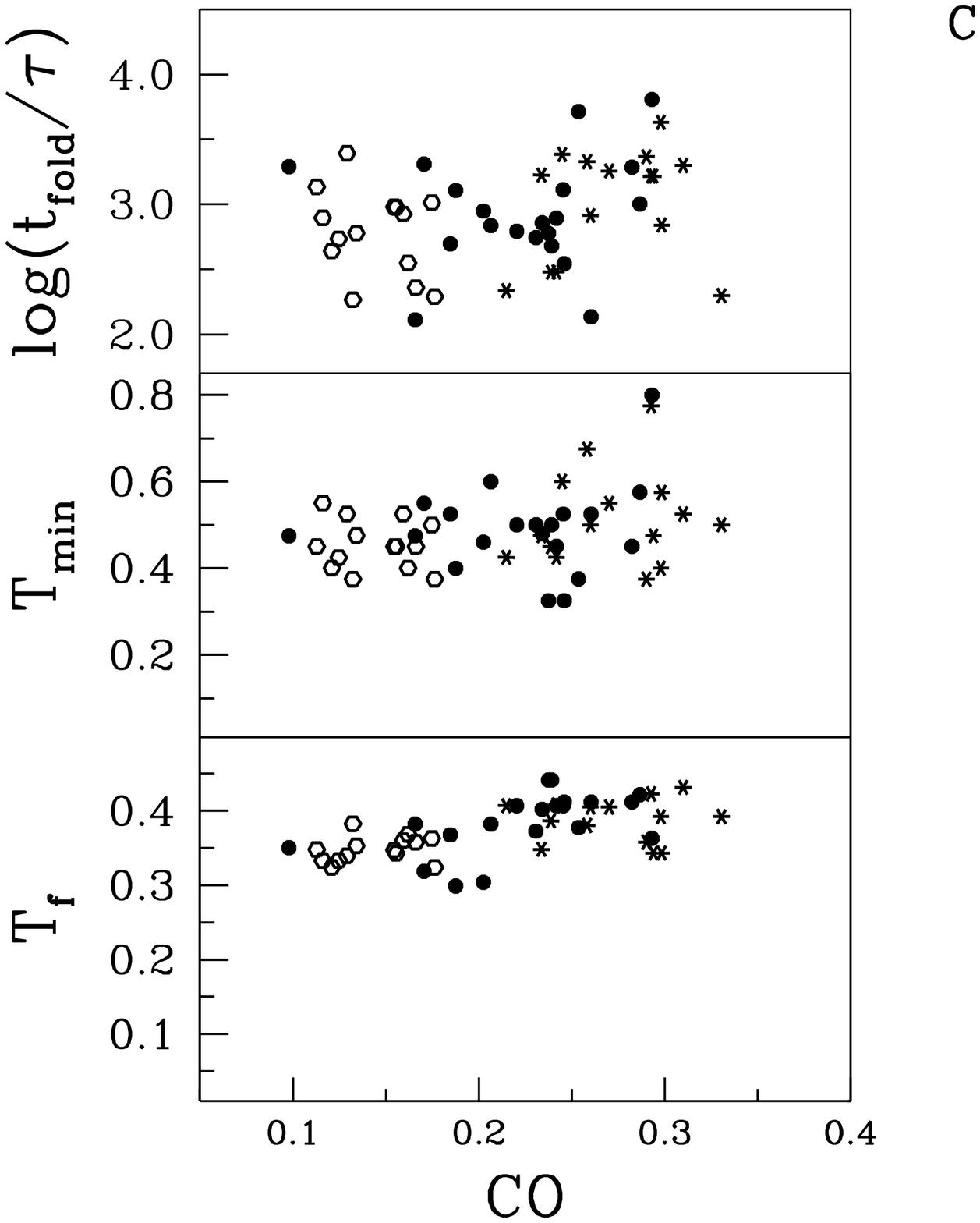}}
\vspace*{3cm}
\caption{ }
\end{figure}

%FIGURE 7
\begin{figure}
\epsfxsize=7in
\centerline{\epsffile{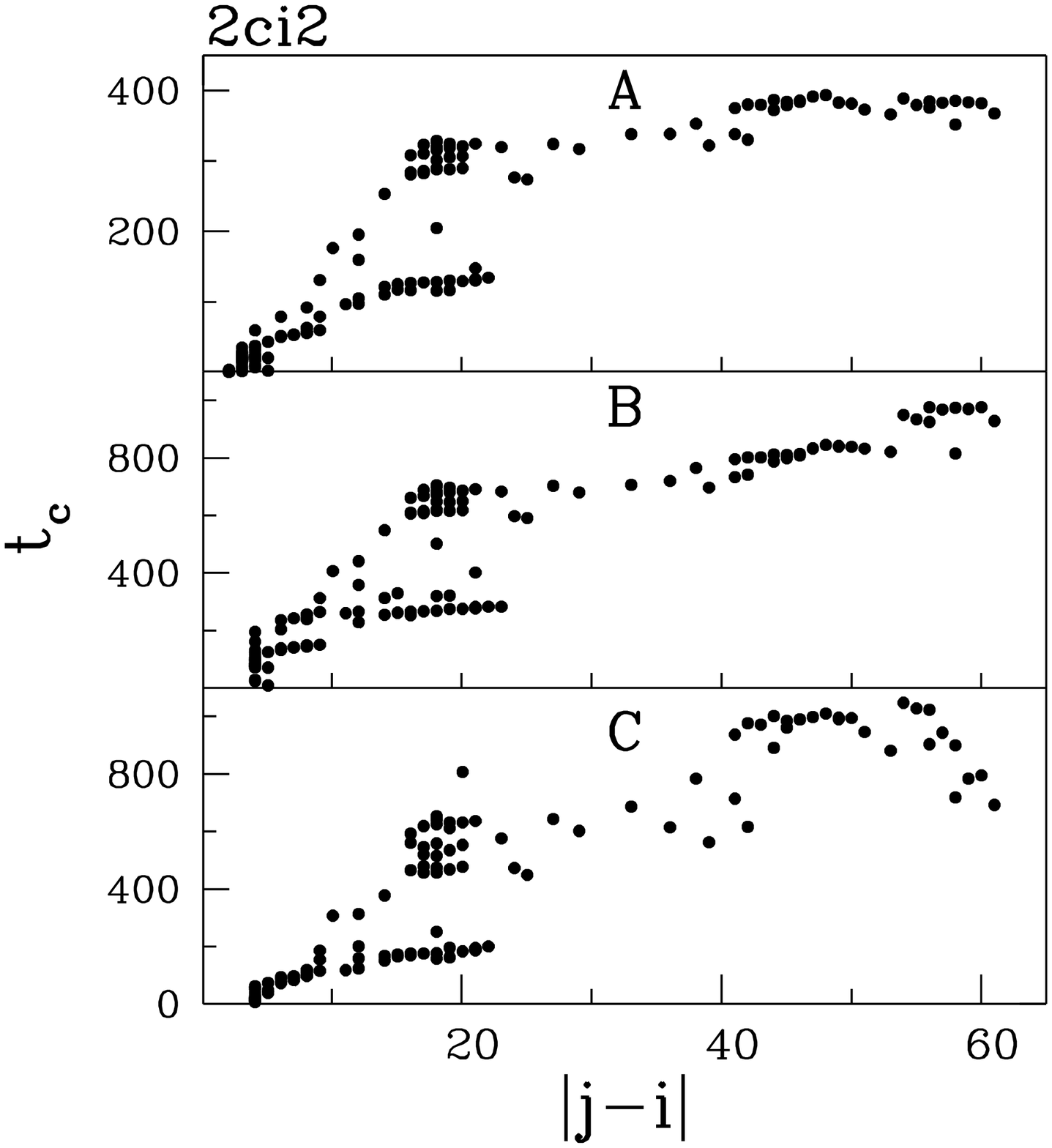}}
\vspace*{3cm}
\caption{ }
\end{figure}

%FIGURE 8
\begin{figure}
\epsfxsize=7in
\centerline{\epsffile{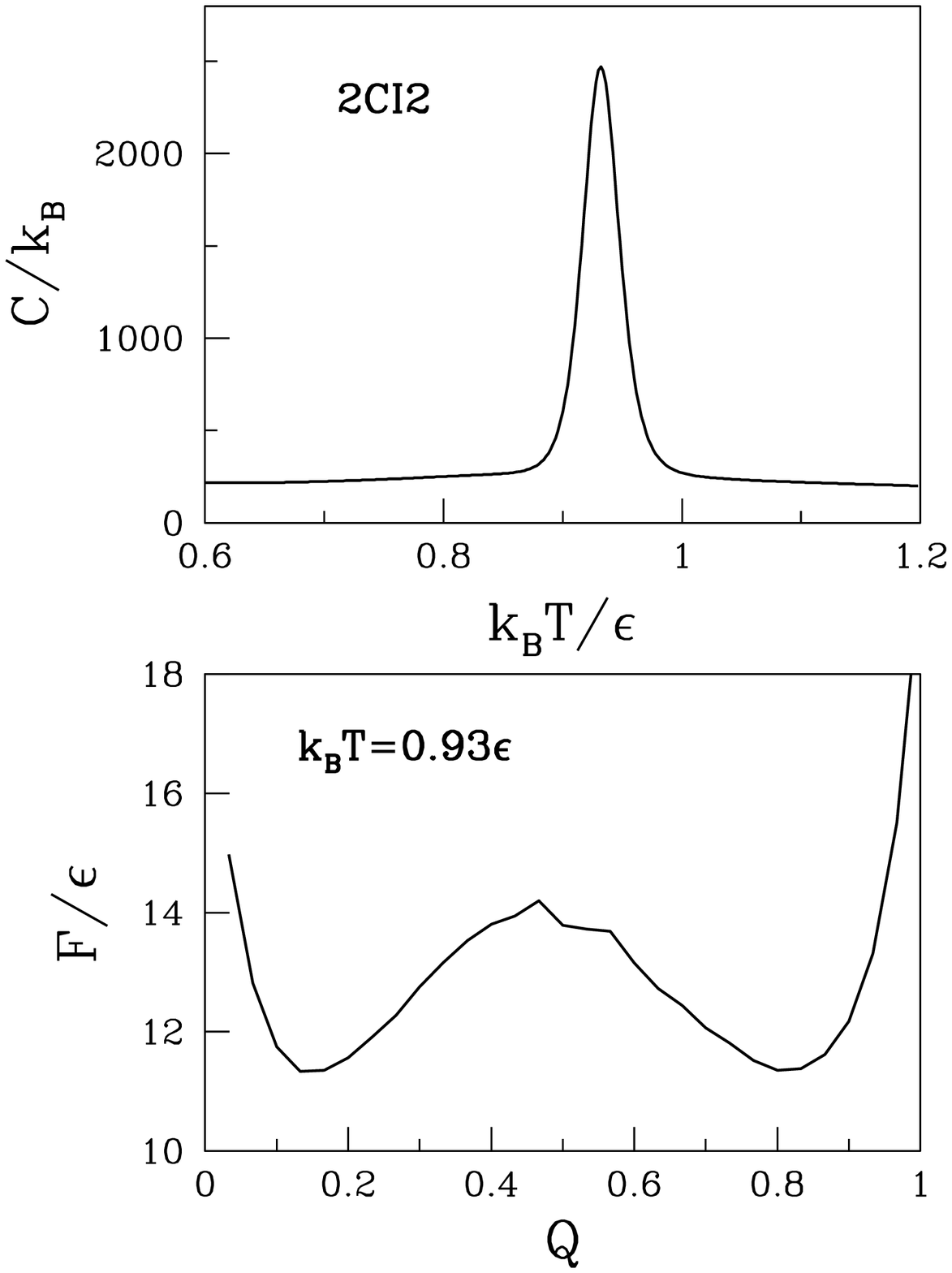}}
\vspace*{3cm}
\caption{ }
\end{figure}

\end{document}